\pdfoutput=1
\documentclass[sigconf]{acmart}
\usepackage{algorithm}
\usepackage{algorithmic}
\usepackage{enumitem}
\usepackage{amsthm}
\usepackage{multirow}
\AtBeginDocument{%
  \providecommand\BibTeX{{%
    \normalfont B\kern-0.5em{\scshape i\kern-0.25em b}\kern-0.8em\TeX}}}

\setcopyright{acmcopyright}
\copyrightyear{2024}
\acmYear{2024}
\acmDOI{10.1145/3589334.3645518}

\acmConference[WWW'24]{International World Wide Web Conference 2024}{May 13--17,
  2024}{Singapore}
\acmBooktitle{WWW'24: International World Wide Web Conference 2024,
 May 13--17, 2024, Singapore} 
\acmPrice{15.00}
\acmISBN{978-1-4503-XXXX-X/18/06}

\begin{document}

%%
%% The "title" command has an optional parameter,
%% allowing the author to define a "short title" to be used in page headers.
\title{Intersectional Two-sided Fairness in Recommendation}

%%
%% The "author" command and its associated commands are used to define
%% the authors and their affiliations.
%% Of note is the shared affiliation of the first two authors, and the
%% "authornote" and "authornotemark" commands
%% used to denote shared contribution to the research.
\author{Yifan Wang}
\orcid{0000-0002-2933-6363}
\affiliation{%
  \institution{Quan Cheng Laboratory \& \\ DCST, Tsinghua University}
  \city{Beijing}
  \postcode{10084}
  \country{China}
}
\email{yf-wang21@mails.tsinghua.edu.cn}

\author{Peijie Sun}
\orcid{0000-0001-9733-0521}
\affiliation{%
  \institution{DCST, Tsinghua University}
  \city{Beijing}
  \postcode{10084}
  \country{China}
}
\email{sun.hfut@gmail.com}

\author{Weizhi Ma$^*$}
\orcid{0000-0001-5604-7527}
\affiliation{%
  \institution{AIR, Tsinghua University}
  \city{Beijing}
  \postcode{10084}
  \country{China}
}
\email{mawz@tsinghua.edu.cn}

\author{Min Zhang}
\authornote{Corresponding authors}
\orcid{0000-0003-3158-1920}
\affiliation{%
  \institution{DCST, Tsinghua University}
  \city{Beijing}
  \postcode{10084}
  \country{China}
}
\email{z-m@tsinghua.edu.cn}

\author{Yuan Zhang}
\orcid{0000-0002-7849-208X}
\affiliation{%
  \institution{Kuaishou Technology}
  \city{Beijing}
  \country{China}
}
\email{yuanz.pku@gmail.com}

\author{Peng Jiang}
\orcid{0000-0002-9266-0780}
\affiliation{%
  \institution{Kuaishou Technology}
  \city{Beijing}
  \country{China}
}
\email{jp2006@139.com}

\author{Shaoping Ma}
\orcid{0000-0002-8762-8268}
\affiliation{%
  \institution{DCST, Tsinghua University}
  \city{Beijing}
  \postcode{10084}
  \country{China}
}
\email{msp@tsinghua.edu.cn}

%%
%% By default, the full list of authors will be used in the page
%% headers. Often, this list is too long, and will overlap
%% other information printed in the page headers. This command allows
%% the author to define a more concise list
%% of authors' names for this purpose.
\renewcommand{\shortauthors}{Wang, et al.}

\newcommand\blfootnote[1]{
    \begingroup
    \renewcommand\thefootnote{}\footnote{#1}
    \addtocounter{footnote}{-1}
    \endgroup
}

% \setcopyright{none}
% \settopmatter{printacmref=false} % Removes citation information below abstract
% \renewcommand\footnotetextcopyrightpermission[1]{} % removes footnote with conference information in first column
% %\pagestyle{plain}

%%
%% The abstract is a short summary of the work to be presented in the
%% article.
\begin{abstract}
  Fairness of recommender systems (RS) has attracted increasing attention recently. 
  Based on the involved stakeholders, the fairness of RS can be divided into user fairness, item fairness, and two-sided fairness which considers both user and item fairness simultaneously.
  However, we argue that the intersectional two-sided unfairness may still exist even if the RS is two-sided fair, which is observed and shown by empirical studies on real-world data in this paper, and has not been well-studied previously.
  To mitigate this problem, we propose a novel approach called \textit{Intersectional Two-sided Fairness Recommendation}~(ITFR). Our method utilizes a sharpness-aware loss to perceive disadvantaged groups, and then uses collaborative loss balance to develop consistent distinguishing abilities for different intersectional groups.
  Additionally, predicted score normalization is leveraged to align positive predicted scores to fairly treat positives in different intersectional groups.
  Extensive experiments and analyses on three public datasets show that our proposed approach effectively alleviates the intersectional two-sided unfairness and consistently outperforms previous state-of-the-art methods.\blfootnote{This work is supported by the Natural Science Foundation of China (Grant No.U21B2026, 62372260), Quan Cheng Laboratory (Grant No. QCLZD202301), the fellowship of China Postdoctoral Science Foundation (No.2022TQ0178) and Kuaishou.}
  
\end{abstract}

%%
%% The code below is generated by the tool at http://dl.acm.org/ccs.cfm.
%% Please copy and paste the code instead of the example below.
%%


%%
%% Keywords. The author(s) should pick words that accurately describe
%% the work being presented. Separate the keywords with commas.

% \received{20  2023}
% \received[revised]{12 March 2009}
% \received[accepted]{5 June 2009}

%%
%% This command processes the author and affiliation and title
%% information and builds the first part of the formatted document.
\maketitle

\begin{figure}[h]
  \centering
  \includegraphics[width=7.8cm]{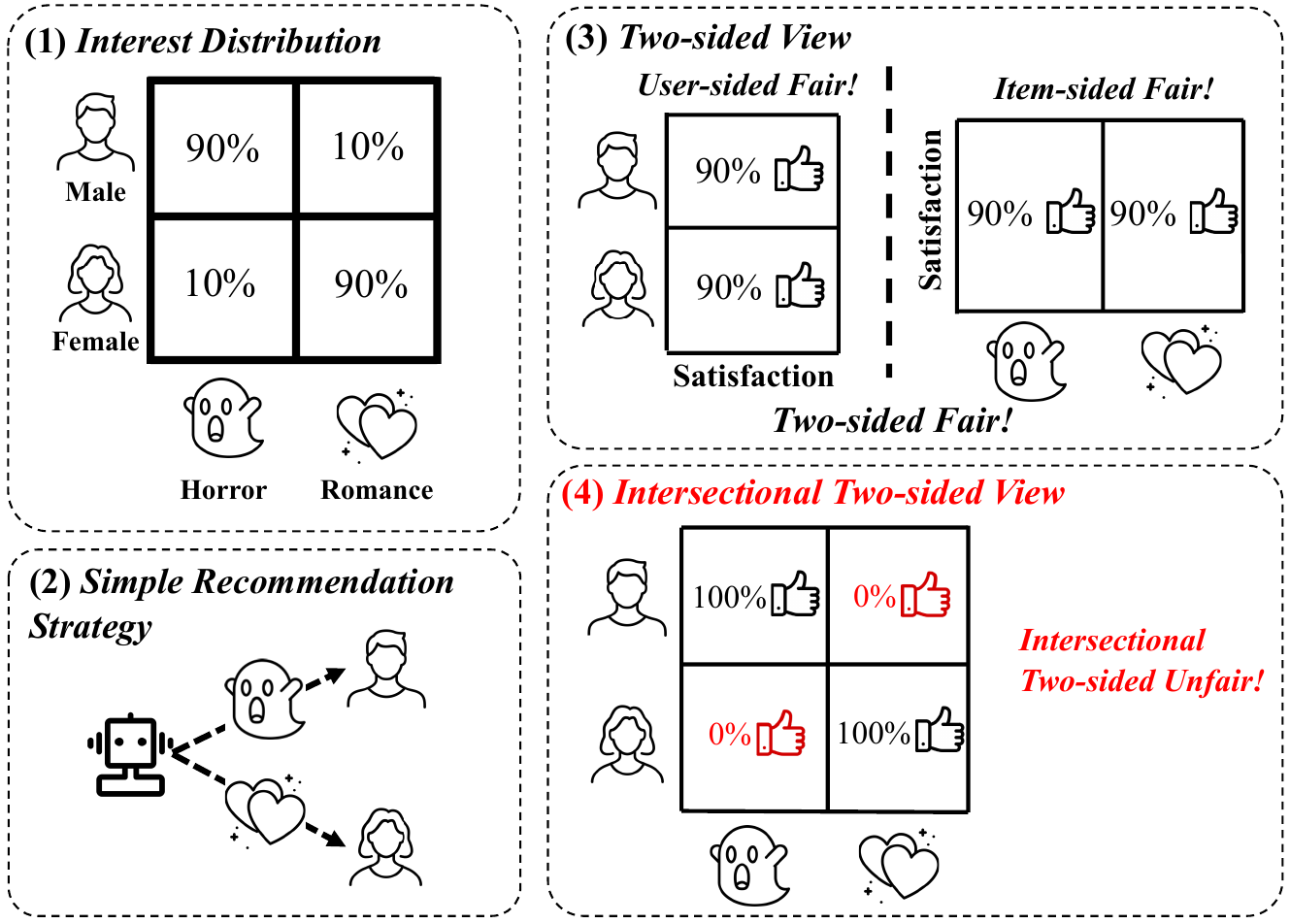}
 \caption{Illustration of intersectional two-sided unfairness. In this toy example, the RS strategy meets two-sided fairness but shows unfair in an intersectional two-sided view. The thumb-up means the recommendation fits the user interest.}
  \label{illustration}
\end{figure}

\section{introduction}
As recommender systems (RS) involve the allocation of social resources, the fairness of RS has attracted increasing attention \cite{ItemFair1, UserFair1, ItemFair2, metricsurvey}. 
Fairness in RS can be divided into three types based on the involved stakeholders: user fairness, item fairness, and two-sided fairness which aims to ensure both user and item fairness concurrently. Presently, user fairness mainly entails consistent recommendation performance across different user groups \cite{UserFair2, UserFair10}, while item fairness primarily focuses on fair exposure \cite{ItemFair1, ItemFair6, ItemFair24} or consistent recommendation performance for different item groups \cite{ItemFair2, ItemFair3}. Two-sided fairness is achieved when both user fairness and item fairness criteria are met simultaneously \cite{TwoSide1, TwoSide2}.

However, we argue that a form of intersectional two-sided unfairness may still exist even when two-sided fairness is achieved, where some intersectional two-sided groups (i.e., the intersection of a user group and an item group, a formal definition can be found in Section 3.1) still face discrimination. 
As illustrated in Fig.\ref{illustration}, we present a toy example. Consider a movie recommendation scenario with 200 users, comprising 100 male and 100 female users, and a movie collection consisting solely of horror and romance genres. Suppose the RS recommends only one movie for each user. Among the male users, 90 prefer horror movies, while the remaining 10 favor romance. Conversely, among the female users, 90 prefer romance movies, and the remaining 10 prefer horror. Now, let us examine a straightforward RS strategy that exclusively recommends horror movies to men and romance movies to women. This recommendation strategy adheres to current two-sided fairness criteria, but the intersectional two-sided groups (Female like Horror movies and Male prefer Romance movies) experience discrimination. While this example is simplified, it can be readily extended to accommodate varying user counts, recommendation lengths, and fair distribution. Such a phenomenon has been observed in the real scenario, which is shown and discussed in Section 3.

Such intersectional two-sided unfairness has manifold harm. For users and items, it hinders part of user interests and item consumers. For the platform, this constrains the development of a diverse user-item ecosystem. Moreover, from a social perspective, such unfairness may reinforce the social polarization issue \cite{ItemFair9}. Note that even if two-sided fairness is met, intersectional two-sided unfairness can still exist and lead to these harms. Hence, addressing intersectional two-sided unfairness is crucial for RS.

To verify the existence of such unfairness, we conduct empirical experiments with real-world data. We find that intersectional unfairness indeed exists and cannot be ignored. Unfortunately, we further observe that current fairness methods cannot effectively mitigate such unfairness, highlighting the importance of designing approaches to address this problem.

To fill this gap, we design a novel method \textit{Intersectional Two-sided Fairness Recommendation} (ITFR) to mitigate such unfairness. Considering that the positives of an intersectional group compete not only with all the negatives but also with positives from other groups in the same recommendation list, we divide the intersectional two-sided fairness into two goals: (i) consistently distinguishing between positives and negatives for different intersectional groups; (ii) fairly treating positives in different intersectional groups. To achieve the first goal, ITFR employs loss balance, as training losses may be a proxy of the distinguishing ability. However, low training losses do not necessarily indicate poor distinguishing ability, and the training losses of various groups can influence each other. Direct reweighting losses based on their size may not be an effective solution. To tackle the first challenge, ITFR incorporates a sharpness-aware loss to improve the alignment between training losses and test performance, thereby enhancing the identification of disadvantaged groups. To address the second challenge, ITFR leverages group collaboration information to learn fair weights for intersectional groups, thereby balancing their sharpness-aware losses. Additionally, to achieve the second goal, predicted score normalization is leveraged to align predicted scores for positives. To demonstrate the effectiveness, we conduct extensive experiments on three public datasets. Experimental results show that our method can effectively alleviate the intersectional two-sided unfairness. Our main contributions can be summarized as follows.

\begin{itemize}[leftmargin=*]
    \item To the best of our knowledge, it is the first work to study the intersectional two-sided fairness in Top-N recommendation. We conduct empirical experiments to show the existence of such unfairness and inadequacy of current fairness methods.
    \item We propose a novel method ITFR to mitigate the intersectional two-sided unfairness, which consists of \textit{sharpness-aware disadvantage group discovery}, \textit{collaborative loss balance}, and \textit{predicted score normalization}.
    \item Extensive experimental results on three public datasets demonstrate that our method can effectively mitigate intersectional two-sided unfairness with similar or even better accuracy.
\end{itemize}
\section{related work}
\subsection{Single-sided Fairness in Recommendation}
Below we introduce the two single-sided fairness in recommendation: user fairness and item fairness.

Current research on user fairness can be roughly divided into two groups: learning fair user representations \cite{UserFair7, UserFair8} and producing fair recommendation outcomes \cite{UserFair2, UserFair3, UserFair5, UserFair6, UserFair9, UserFair10}. The former is related to process fairness, while the latter focuses on the fairness of recommendation performance received by different users, which has attracted more attention as it is more related to user satisfaction. These fairness methods on outcome fairness can be roughly grouped into three categories: (i) fairness regularization \cite{UserFair3,UserFair5,UserFair9}. (ii) distributionally robust optimization \cite{UserFair10, UserFair11}. (iii) re-ranking \cite{UserFair2, UserFair6}.

Most existing item fairness studies can be divided into exposure-based fairness (or treatment-based fairness) and performance-based fairness (or impact-based fairness). Exposure-based item fairness focuses on allocating fair exposure to each item \cite{ItemFair5, ItemFair14, ItemFair19, ItemFair25}. Most of them propose integer programming-based re-ranking methods \cite{ItemFair1, ItemFair6, ItemFair7, ItemFair8, ItemFair10, ItemFair11, ItemFair12, ItemFair15, ItemFair16, ItemFair17, ItemFair22}. Unlike exposure-based item fairness, performance-based item fairness focuses on whether different item groups have consistent recommendation performance (e.g., recall of the positives), which is related to users' true preferences. The fairness methods on performance-based fairness can be coarsely divided into three categories: (i) fairness regularization \cite{ItemFair4, ItemFair13}. (ii) adversarial learning \cite{ItemFair2}. (iii) fairly negative sampling \cite{ItemFair3}.

The above methods only enhance single-sided fairness. However, since the RS is a typical two-sided platform, it is important to ensure both user and item fairness.

\subsection{Two-sided Fairness in Recommendation}
Current work \cite{TwoSide1,TwoSide2,TwoSide3,TwoSide4,TwoSide5,Twoside6} on two-sided fairness in Top-N recommendation is aimed to ensure user and item fairness simultaneously. Specifically, most studies focus on ensuring performance-based user fairness (i.e., different users receive consistent recommendation performance) and exposure-based item fairness, except for \cite{Twoside6} focusing on purely exposure fairness in a stochastic ranking scenario. Most work \cite{TwoSide2,TwoSide3,TwoSide4,TwoSide5} designs fair re-ranking methods to achieve this goal as the allocation of exposure is more feasible in the re-ranking stage, while \cite{TwoSide1} propose a multi-objective optimization approach in the ranking stage. Unlike these studies, we focus on performance-based fairness both for users and items, which will be further explained in Section 3.1. A similar study is \cite{Joint1}, which focuses on the marketing bias in the rating prediction task and also involves intersectional groups. However, the intersectional unfairness in this paper may not be due to the marketing bias, and their method is not designed for Top-N recommendation.

Unlike current work, we argue that ensuring user and item fairness simultaneously is insufficient. This paper aims to alleviate the intersectional two-sided unfairness in Top-N recommendation, which current fairness methods may overlook.

\subsection{Intersectional Fairness in Machine Learning}
There have been several studies \cite{InterML1, InterML2, InterML3, InterML4, InterML5, InterML6, InterML7, InterML8} on intersectional fairness in machine learning (ML), which focus on the intersectional groups of different attributes, such as race \& gender (e.g., black females). These studies argue that when multiple fairness-aware attributes exist, each intersectional group should be treated fairly. Multiple attribute divisions may lead to more sparse and unbalanced subgroups, which is the concern of these studies.

Different from these studies on intersectional fairness in ML, we focus on the intersectional two-sided fairness in the Top-N recommendation, which has a key difference: the two-sided group makes it more challenging than the single-sided group, i.e., the single-sided fairness methods and current two-sided fairness methods ignoring the intersectional groups is not designed to mitigate such intersectional two-sided unfairness and cannot be used on intersectional two-sided groups directly. However, the intersectional single-sided fairness (e.g., the unfair recommendation performance of black females) may be effectively alleviated by current adequate single-sided fairness methods, as we can just treat the intersectional single-sided groups as a new single-sided group division.
%\section{Preliminaries}
\section{Problem Definition and \\ Empirical Study}
\subsection{Problem Definition}
Suppose there are $n$ users $\mathcal{U} = \{u_1,...,u_n\}$ and $m$ items $\mathcal{V} = \{v_1,...,v_m\}$. The collected user feedback can be represented by $\mathcal{Y} \in \{0,1\}^{n \times m}$, where $y_{ui}$ denotes whether the user $u$ has interacted with the item $i$. The whole positives are $\mathcal{D} = \{(u,i)| y_{u,i} = 1\}$. Ideally, there exists an unobserved matrix $\mathcal{R} \in \{0,1\}^{n \times m}$, where $r_{ui}$ represents whether a user $u$ will interact with an item $i$. The top-N recommendation task is to recommend a list of $N$ uninteracted items to each user $u$.

In this paper, we focus on group-level fairness. Suppose the users and items are divided into $P$ and $Q$ disjoint groups by some predefined attributes, respectively. As each interaction belongs to a user group and an item group simultaneously, we are particularly interested in the following intersectional two-sided groups.

\begin{definition}[Intersectional Two-sided Group] Given $P$ disjoint user groups $\{\mathcal{U}_1,...,\mathcal{U}_P\}$ and $Q$ disjoint item groups $\{\mathcal{V}_1,...,\mathcal{V}_Q\}$, the potential interactions $\mathcal{R}$ can be divided into $P \times Q$ intersectional two-sided groups, i.e., $\{g_{m,n}\coloneqq(\mathcal{U}_m, \mathcal{V}_n)| 1\le m\le P, 1\le n\le Q\}$. 
\end{definition}

To study the fairness of these intersectional two-sided groups, we further define the utility of these groups. Specifically, the utility of an intersectional two-sided group should reflect the received recommendation performance of potential interactions in this group. Formally, let $\mathcal{U}(i,j)$ denote all users in the $i$-th user group who are interested in at least one uninteracted item in the $j$-th item group $\mathcal{V}_j$. The utility of the intersectional two-sided group $g_{i,j}$ is defined as the average utility for these potential interests:
\begin{equation}
\label{utility_definition}
    \text{ITG\_Utility}(i, j)@K = \frac{1}{|\mathcal{U}(i,j)|} \sum_{u \in \mathcal{U}(i,j)} \text{utility}(u, j)@K
\end{equation}
where $\text{utility}(u, j)@K$ can be some metrics for recommendation performance. Without loss of generality, we follow \cite{ItemFair2, ItemFair3} and use a recall-based metric, i.e., $\text{utility}(u, j)@K = \frac{|\{i| i \in l_u \& r_{u,i} = 1 \& i \in \mathcal{V}_j\}|}{|\{i| y_{u,i} = 0 \& r_{u,i} = 1 \& i \in \mathcal{V}_j\}|}$, here $l_u$ is the top-K recommendation list for user $u$.\footnote{Note that we ignore users not interested in the $j$-th item group, as the utility for these users is always zero and meaningless.}

Based on the utility definition, intersectional two-sided fairness aims to provide similar utilities for different groups. Intuitively, it seeks independence of the utilities of different groups and their two-sided group information. Formally, we can define the \textbf{$\epsilon-$Intersectional Two-sided Fairness}.
\begin{definition}[$\epsilon-$Intersectional Two-sided Fairness]
A recommender system satisfies $\epsilon-$intersectional two-sided fairness when, for any different two intersectional two-sided groups $g_{u_a, v_a}$ and $g_{u_b, v_b}$, we have $\lvert \text{ITG\_Utility}(u_a, v_a) - \text{ITG\_Utility}(u_b, v_b) \rvert \le \epsilon$, where $\epsilon$ is a sufficiently small positive number.\footnote{For simplification, we omit the $\epsilon$ in the remaining part of the paper.}
\end{definition}
The reason to choose such a performance-based utility definition instead of an exposure-based utility definition (e.g., the received exposure for intersectional groups) is that the latter does not consider user preferences. Specifically, for user fairness, the exposure-based utility is inconsistent with current user fairness that focuses on recommendation performance related to user preferences \cite{UserFair10, UserFair6}. For item fairness, it is also crucial to consider user preferences \cite{ItemFair3}. Only exposure-based item fairness without performance-based fairness might cause some item groups to receive low recommendation quality, i.e., recommended to the users uninterested in them \cite{fairsurvey}. 

\subsection{Existence of Intersectional Unfairness}
Below we investigate whether such unfairness exists in real datasets. For brevity, we only use a classic dataset Movielens-1M (ML1M) to conduct our empirical experiments. 
Here we only consider the binary group setting, and in subsequent experimental sections, we show results for more than two groups. 
We use gender to divide users (Male v.s. Female) and movie genres to divide items (here we take `Children's v.s. Horror' as an example). The processed dataset ML1M-2 contains 4,403 users, 568 items, and 144,420 interactions. We randomly divide all interactions into training, validation, and test sets in the ratio of 7:1:2. We run the classic BPR \cite{BPR} algorithm and repeat it five times, and the results are shown in Table \ref{bpr_binary}, where URecall@20 is the average Recall@20 for the user group, and IRecall@20 is the recall at the item group level \cite{ItemFair2}.

\begin{table}[h]\small
\centering
\caption{Results of BPR (ML1M-2). \textit{Italic} for the bottom two intersectional groups and the worst single-sided group.}
\label{bpr_binary}
\begin{tabular}{ll|cc|l}
\toprule
\multicolumn{2}{l|}{\multirow{2}{*}{\textbf{ITG$\_$Utility@20}}}                 & \multicolumn{2}{c|}{\textbf{User}} & \multirow{2}{*}{\textbf{IRecall@20}} \\ \cline{3-4}
\multicolumn{2}{l|}{}                                         &\textbf{Female}         & \textbf{Male}          &                             \\ \midrule
\multicolumn{1}{l|}{\multirow{2}{*}{\textbf{Item}}} & \textbf{Children's} &     0.5215       &      \textit{0.4669}         &  0.4440                      \\
\multicolumn{1}{l|}{}                            & \textbf{Horror}     &   \textit{0.4125}         &    0.4814           &   \textit{0.4101}                          \\ \midrule
\multicolumn{2}{l|}{\textbf{URecall@20}}                               &      0.5070      &     \textit{0.5018}             &   -                          \\ \bottomrule
\end{tabular}%
\end{table}

From the single-sided fairness perspective, we can find that `Children's' gets a significantly better performance in terms of item fairness, while male and female users get very similar performance without significant differences. Existing two-sided fairness only requires single-sided fairness for both users and items. Therefore, the model should improve the performance of `Horror' in terms of item fairness while keeping the current fair status on the user side.

However, the story is different from the intersectional two-sided perspective. As shown in Table 1, these four intersectional two-sided groups receive inconsistent recommendation quality, with (Male \& Children's) and (Female \& Horror) receiving worse performance. The best group (Female \& Children's) has an about 26\% performance gap compared to the worst group (Female \& Horror), which indicates that intersectional two-sided unfairness indeed exists. Note that the two best intersectional groups are on the diagonal, which is consistent with Fig.\ref{illustration}.

\subsection{Do Current Methods Help?}
Next, we investigate the effectiveness of current fairness methods. We use two advanced single-sided fairness methods: FairNeg \cite{ItemFair3} for item fairness and StreamDRO \cite{UserFair10} for user fairness, which both are focused on performance-based fairness and applied in the ranking stage. Although there is no performance-based two-sided fairness method, for experimental completeness, we use an in-processing two-sided fairness method MultiFR \cite{TwoSide1}, focusing on performance-based user fairness and exposure-based item fairness.

\begin{table}[h]\small
\centering
\caption{Results of FairNeg (item fairness) on the ML1M-2 dataset. \textit{Italic} for the bottom two intersectional groups and the worst single-sided group. The (↑/↓) means better or worse recommendations compared with BPR in Table \ref{bpr_binary}.}
\label{fairneg_binary}
\begin{tabular}{ll|cc|l}
\toprule
\multicolumn{2}{l|}{\multirow{2}{*}{\textbf{ITG$\_$Utility@20}}}                 & \multicolumn{2}{c|}{\textbf{User}} & \multirow{2}{*}{\textbf{IRecall@20}} \\ \cline{3-4}
\multicolumn{2}{l|}{}                                         & \textbf{Female}          & \textbf{Male}          &                             \\ \midrule
\multicolumn{1}{l|}{\multirow{2}{*}{\textbf{Item}}} & \textbf{Children's} &    0.5042(↓)      &     \textit{0.4490}(↓)          &  0.4281(↓)                            \\
\multicolumn{1}{l|}{}                            & \textbf{Horror}     &     \textit{0.4258}(↑)        &  0.4956(↑)            &    \textit{0.4266}(↑)                         \\ \bottomrule
\end{tabular}%
\end{table}

\begin{table}[h]\small
\centering
\caption{Results of StreamDRO (user fairness) on the ML1M-2 dataset. The notations are similar to Table \ref{fairneg_binary}.}
\label{stream_binary}
\begin{tabular}{ll|cc}
\toprule
\multicolumn{2}{l|}{\multirow{2}{*}{\textbf{ITG$\_$Utility@20}}}                 & \multicolumn{2}{c}{\textbf{User}} \\ \cline{3-4} 
\multicolumn{2}{l|}{}                                         & \textbf{Female}          & \textbf{Male}         \\ \midrule
\multicolumn{1}{l|}{\multirow{2}{*}{\textbf{Item}}} & \textbf{Children's} &     0.5202(↓)       &   \textit{0.4696}(↑)             \\
\multicolumn{1}{l|}{}                            & \textbf{Horror}     &      \textit{0.4121}(↓)     &   0.4816(↑)            \\ \midrule
\multicolumn{2}{l|}{\textbf{URecall@20}}                               &   0.5066(↓)       &   \textit{0.5031}(↑)          \\ \bottomrule
\end{tabular}%
\end{table}

As shown in Tables \ref{fairneg_binary} and \ref{stream_binary}, current single-sided fairness methods indeed improve targeted single-sided fairness. However, they do not improve intersectional two-sided fairness very well. For item fairness, FairNeg indeed narrows the overall gap between item groups, improving item fairness in the single-sided view. However, in the intersectional view, it improves the performance for all the Horror groups, leading to better performance for some advantaged groups (Male \& Horror) and worse performance for some disadvantaged groups (Male \& Children's). For user fairness, a similar phenomenon can be found in Table \ref{stream_binary}, where some advantaged groups (Male \& Horror) receive better recommendations and some disadvantaged groups (Female \& Horror) receive worse recommendations.

\begin{table}[h]\small
\centering
\caption{Results of MultiFR (two-sided fairness) on the ML1M-2 dataset. The notations are similar to Table \ref{fairneg_binary}.}
\label{multifr_binary}
\begin{tabular}{ll|cc|l}
\toprule
\multicolumn{2}{l|}{\multirow{2}{*}{\textbf{ITG$\_$Utility@20}}}                 & \multicolumn{2}{c|}{\textbf{User}} & \multirow{2}{*}{\textbf{IRecall@20}} \\ \cline{3-4}
\multicolumn{2}{l|}{}                                         &\textbf{Female}         & \textbf{Male}          &                             \\ \midrule
\multicolumn{1}{l|}{\multirow{2}{*}{\textbf{Item}}} & \textbf{Children's} &     0.5151(↓)       &      \textit{0.4671}(↑)        & 0.4394(↓)                       \\
\multicolumn{1}{l|}{}                            & \textbf{Horror}     &   \textit{0.4072}(↓)         &    0.4809(↓)           &  \textit{0.4086}(↓)                           \\ \midrule
\multicolumn{2}{l|}{\textbf{URecall@20}}                               &     0.5019(↓)       &       \textit{0.5016}(↓)           &   -                          \\ \bottomrule
\end{tabular}%
\end{table}

The results for two-sided fairness methods are shown in Table \ref{multifr_binary}. We can find that it indeed narrows the gap between different user groups, but does not effectively improve performance-based item fairness as it considers exposure-based fairness. It can also be found that the worst intersectional group (Female \& Horror) in Table \ref{bpr_binary} receives worse recommendations.

As current methods cannot effectively mitigate such unfairness, it is important to design an effective fairness method for improving intersectional two-sided fairness.
\section{Intersectional Two-sided Fairness Recommendation}
\begin{figure*}[t]
  \centering
  \includegraphics[width=16cm]{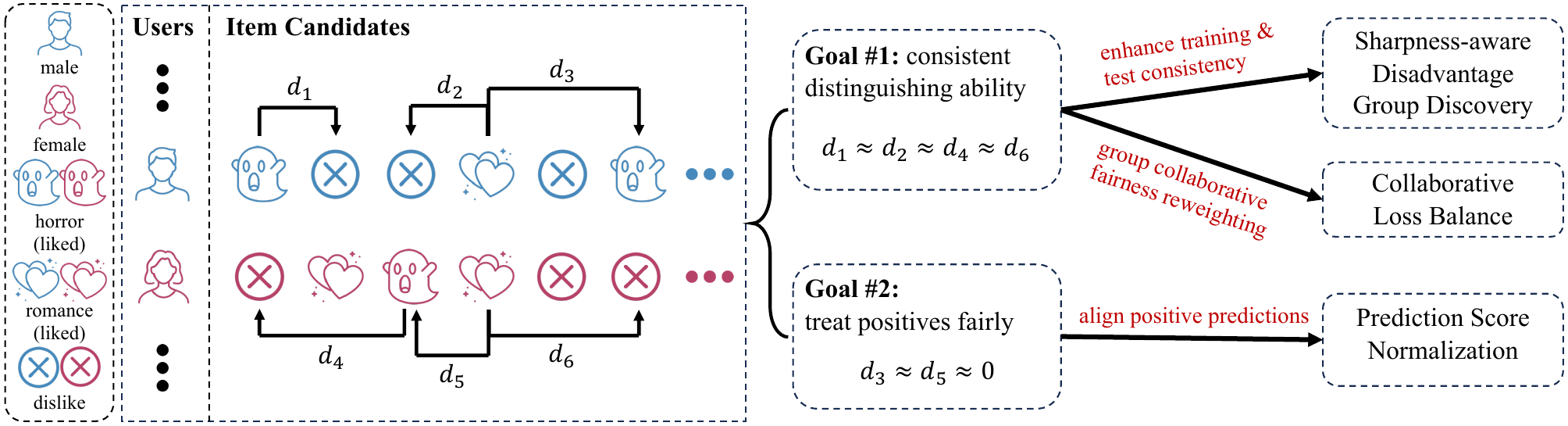}
 \caption{Illustration for our method. $d$ denotes the difference in predicted scores between two interactions.}
  \label{method_illustration}
\end{figure*}

\subsection{Overview}
The utility of an intersectional group is determined by the rank of the positives in this group in the recommendation lists. These positives compete with two kinds of samples: all the negatives and other positives from distinct groups in the recommendation list. Thus, we divide the intersectional two-sided fairness into two goals to balance these competitions separately. As shown in Fig.\ref{method_illustration}, (i) the RS should consistently distinguish between positives and negatives for different intersectional groups; (ii) the RS should treat positives in different intersectional groups fairly to ensure that no positives in a certain group have systematically low predicted scores.

To achieve the above two goals, we propose a method \textit{Intersectional Two-sided Fairness Recommendation} (ITFR), which consists of three components: \textit{sharpness-aware disadvantage group discovery}, \textit{collaborative loss balance}, and \textit{predicted score normalization}. The purpose of the first two components is to balance the training losses between different intersectional groups, which reflects the ability to distinguish between positives and negatives, corresponding to the first goal. Nevertheless, low training losses do not necessarily indicate poor test performance, and different intersectional groups are related to each other. Direct reweighting losses based on their size may not be an effective solution. To tackle the first challenge, we introduce \textit{sharpness-aware disadvantage group discovery} to enhance the consistency of training losses and test performance. To address the second challenge, we leverage the group collaboration information to learn fair weights for these intersectional groups, i.e., \textit{collaborative loss balance}.

However, only controlling the training loss may not meet the second goal. Even if the training loss is similar between different intersectional groups, the predicted score for positive samples in different intersectional groups may be systematically biased as the commonly used recommendation loss (e.g., BPR) only optimizes the distance between positives and negatives and does not constrain the absolute value of the predicted scores. Therefore, the third component, \textit{predicted score normalization}, is applied to achieve the second goal by aligning positive predictions. Next, we elaborate on our method from the above three components respectively.

\subsection{Sharpness-aware Disadvantage Group Discovery}
Let us first consider the first goal, i.e., to fairly distinguish between positives and negatives for different intersectional groups. First, we need to perceive those intersectional groups with poor distinguishing ability on the test data. Since test data is not available, the intuitive idea is to treat the training loss as a proxy for distinguishing ability on the test data, as they mostly reflect the ability of RS to distinguish positives and negatives, and can be optimized by the commonly used gradient descent. Higher training loss is likely to represent poorer distinguishing ability.

Below we formalize the training loss of the intersectional groups. Take the most commonly used recommendation loss BPR \cite{BPR} as an example. The BPR loss $\mathcal{L}_{p,q}$ for an intersectional group $g_{p,q}$ is defined as follows:
\begin{equation}
    \mathcal{L}_{p,q}(\mathcal{D}; \theta) := \frac{1}{\left| \tilde{\mathcal{D}}_{p,q} \right|} \sum_{(u,i,j) \in \tilde{\mathcal{D}}_{p,q}} \text{BPR}(u,i,j)
\end{equation}
Here $\tilde{\mathcal{D}}_{p,q} = \{(u,i,j)| u \in \mathcal{U}_p, i \in \mathcal{V}_q, j \in \mathcal{V}, y_{u,i} = 1, y_{u,j} = 0 \}$, $\text{BPR}(\cdot)$ is the BPR loss for a triple pair.

Moreover, in addition to the value of training losses, the geometric properties (e.g., sharpness) of the loss around the parameters $\theta$ also impact the test performance \cite{sharpness}. Considering training losses on only a single point $\theta$ is vulnerable to random perturbations if the loss curve is sharp, which may lead to an ineffective detection of discriminated groups. To alleviate this, inspired by related work in machine learning \cite{sharpness}, we use the worst training loss within a bounded region of the current parameters $\theta$ as a proxy for the model's distinguishing ability. Formally, the worst loss of the model parameter $\theta$ in the $\rho$-region (i.e., $\left\{ \theta + \epsilon \mid \| \epsilon \| \le \rho \right\}$) for the group $g_{p,q}$ can be defined as:
\begin{equation}
\label{sharpness_cal}
    \hat{\mathcal{L}}_{p,q}(\mathcal{D}, \theta, \rho) := \max_{ \| \epsilon \| \le \rho} \mathcal{L}_{p,q}(\mathcal{D}; \theta + \epsilon)
\end{equation}
Compared with the original loss $\mathcal{L}_{p,q}(\mathcal{D}; \theta)$, the loss $\hat{\mathcal{L}}_{p,q}$ considers the sharpness of the original loss around the parameters $\theta$ since a larger sharpness will lead to a larger difference between the original loss and the worst loss.

\textbf{Practice Detail.} The above definition is not feasible in practice as solving Eq.(\ref{sharpness_cal}) is time-costing. Following \cite{sharpness}, we use single-step gradient ascent to approximate the worst loss. The corresponding solution for Eq.(\ref{sharpness_cal}) is $\theta_{p,q}^* = \theta + \epsilon_{p,q}^*$, where $\epsilon_{p,q}^* = \rho \frac{\nabla_\theta \mathcal{L}_{p,q}(\mathcal{D};\theta)}{\|\nabla_\theta \mathcal{L}_{p,q}(\mathcal{D};\theta)\|}$.

\subsection{Collaborative Loss Balance}
As the sharpness-aware loss reflects the distinguishing ability, the next question is how to balance the sharpness-aware loss $\hat{\mathcal{L}}_{p,q}(\mathcal{D}, \theta, \rho)$ between different intersectional groups. An intuitive idea is reweighting, i.e., assigning higher training weights to groups with higher losses. The group distributionally robust optimization (GroupDRO) \cite{GroupDRO} in machine learning can be leveraged to achieve this goal. Specifically, GroupDRO will assign a weight $w_{i}$ for the $i$-th group, and calculate the total loss as $\mathcal{L}_{GroupDRO} = \sum_{i} w_{i} \mathcal{L}_{i}(\theta)$, where $\mathcal{L}_{i}(\theta)$ is the average training loss of the $i$-th group, and $w_i$ is updated at each batch by:
\begin{equation}
\label{groupdroalg}
    w_{i} \leftarrow \frac{w_{i} \cdot \text{exp}(\eta \cdot \mathcal{L}_{i}(\theta))}{ \sum_{j} w_{j} \cdot \text{exp}(\eta \cdot \mathcal{L}_{j}(\theta))}
\end{equation}
where $\eta$ is a hyperparameter. We can adopt the GroupDRO for intersectional two-sided fairness by replacing $w_i$ with $w_{p,q}$ and $\mathcal{L}_{i}(\theta)$ with $\mathcal{L}_{p,q}(\mathcal{D};\theta)$. Note that the original GroupDRO does not consider the sharpness of losses. To capture the sharpness, we can directly replace $\mathcal{L}_{p,q}(\mathcal{D};\theta)$ with $\hat{\mathcal{L}}_{p,q}(\mathcal{D}, \theta, \rho)$.

However, the above method has a drawback as it ignores the collaboration between different intersectional groups, which is important for RS. Typically, there is much collaborative information in users' interactions, and different intersectional groups are related to each other as they may share similar users and items. Therefore, optimizing the loss of one group can also strongly impact the losses of other groups. The current weight $w_{p,q}$ only considers the current group loss $\hat{\mathcal{L}}_{p,q}$ and does not consider the influence between different groups, which may lead to suboptimal performance.

To model the relationship between different intersectional groups, we define the contribution of a group $g_{p, q}$ to $g_{a, b}$ as the change of sharpness-aware loss of $g_{a,b}$ after updating $\theta$ using the sharpness-aware loss of $g_{p,q}$, where $\nabla_\theta \hat{\mathcal{L}}_{p,q}$ is the gradients of $\hat{\mathcal{L}}_{p,q}(\mathcal{D}, \theta, \rho)$:
\begin{equation}\label{pair_contrib}
    \begin{aligned}
        \mathcal{C}(g_{p,q} \rightarrow g_{a,b}) &:= \hat{\mathcal{L}}_{a,b}(\mathcal{D}, \theta - \alpha \nabla_\theta \hat{\mathcal{L}}_{p,q}, \rho) - \hat{\mathcal{L}}_{a,b}(\mathcal{D}, \theta, \rho)\\
        &\approx \mathcal{L}_{a,b}(\mathcal{D};\theta_{a,b}^*- \alpha \nabla_\theta \hat{\mathcal{L}}_{p,q}) - \hat{\mathcal{L}}_{a,b}(\mathcal{D}, \theta, \rho)
    \end{aligned}
\end{equation}

Furthermore, the total contribution of an intersectional group is defined as a weighted sum of its contributions to all groups:
\begin{equation}
\label{all_contrib}
    \mathcal{C}(g_{p,q}) = \sum_{a = 1}^P \sum_{b = 1}^Q \beta_{a,b} \mathcal{C}(g_{p,q} \rightarrow g_{a,b}),\text{where} \; \beta_{a,b} = \frac{ (\mathcal{L}_{a,b})^\gamma}{\sum_{p,q} (\mathcal{L}_{p,q})^\gamma}
\end{equation}
Here we introduce the group weight $\beta_{a,b}$ because drops in larger losses are more valuable. $\gamma$ is a hyperparameter, and a larger $\gamma$ means we pay more attention to disadvantaged groups.

Given the total contribution $\mathcal{C}(g_{p,q})$ of each group $g_{p,q}$, we can calculate the group weight $w_{p,q}$ following Eq.(\ref{groupdroalg}):
\begin{equation}
\label{w_update}
    w_{p,q} \leftarrow \frac{w_{p,q} \cdot \text{exp}(\eta \cdot \mathcal{C}(g_{p,q}))}{ \sum_{a,b} w_{a,b} \cdot {exp}(\eta \cdot \mathcal{C}(g_{a,b}))}
\end{equation}

The final collaborative balanced loss is $\mathcal{L}_{clb} = \sum_{p,q} w_{p,q} \cdot \hat{\mathcal{L}}_{p,q}$. The total procedure for Goal \#1 can be found in Appendix A.1.

\textbf{Practice Detail.} In practice, Eq.(\ref{pair_contrib}) will introduce a high computational cost. Following \cite{CGD}, we use the first-order Taylor approximation and get $\mathcal{C}(g_{p,q} \rightarrow g_{a,b}) \approx \alpha \nabla_\theta \hat{\mathcal{L}}_{p,q}^T \nabla_\theta \hat{\mathcal{L}}_{a,b}$, the $\nabla_\theta \hat{\mathcal{L}}_{p,q}$ here is further approximated by $\sqrt{\hat{\mathcal{L}}_{p,q}} \frac{\nabla_\theta \hat{\mathcal{L}}_{p,q}}{\| \nabla_\theta \hat{\mathcal{L}}_{p,q} \|}$ to obtain a stable optimization, and the $\nabla_\theta \hat{\mathcal{L}}_{a,b}$ is approximated similarly. Besides, as the parameters are not shared between different users and items in common ID-based RSs, and two groups within a batch may not have overlapped users and items, the $\nabla_\theta \hat{\mathcal{L}}_{p,q}^T \nabla_\theta \hat{\mathcal{L}}_{a,b}$ will be zero as the gradients of two intersectional groups have no overlap in each batch. To alleviate this problem, we use the cumulative gradients of the last epoch as the approximation of $\nabla_\theta \hat{\mathcal{L}}_{a,b}$.

\subsection{Predicted Score Normalization}
Although the proposed loss balance method can improve the first goal, it may not necessarily satisfy the second goal, i.e., to fairly treat positives in different intersectional groups. This is because the commonly used BPR loss only optimizes the distance between positives and negatives. Even if the distances between positives and negatives are the same across intersectional groups, there may still be systematic unfairness in their predicted scores for positives. Note that the competition between positives occurs only between items and that the predicted scores of positive samples are not comparable across users, so this issue may have a more significant impact on item fairness than user fairness.

As directly controlling predicted scores may result in a large accuracy loss \cite{ItemFair2}, we leverage an indirect approach here to alleviate this problem. Note that the proposed loss balance method enhances the similarity of distances between positives and negatives across different intersectional groups. If the range of predicted scores is bounded, then the systematic bias between different intersectional groups may be mitigated, as this bias is restricted to a certain bound rather than over the real number domain. Thus, given user embedding $u$ and item embedding $v$, we bound the commonly used inner product predicted scores $\hat{y}_{u,v} = u^Tv$ to $(-\tau, \tau)$, formally, in an embedding normalization manner:
\begin{equation}
\hat{y}_{u,v} = \tau \cdot \frac{u^Tv}{\|u \| \|v \|}.
\end{equation}

There could be other ways to normalize the predicted scores, e.g., $\hat{y}_{u,v} = \tau \cdot \text{sigmoid}(u^Tv)$. However, the embedding normalization manner has its unique advantages: (i) the normalization of user embeddings makes training losses more comparable across users, given that the magnitude of user embeddings influences the training losses but does not affect the recommendation lists. (ii) the normalization of item embeddings may partially alleviate the popularity bias \cite{adaptau}, which may be one of the reasons for the predicted score inconsistency between different item groups.

\section{experiments}
\subsection{Datasets and Settings}
\subsubsection{Datasets}
Experiments are conducted on three public datasets: \textbf{Movielens}-1M\footnote{https://grouplens.org/datasets/movielens/1m/}, \textbf{Tenrec}-QBA\footnote{https://static.qblv.qq.com/qblv/h5/algo-frontend/tenrec\_dataset.html/} \cite{Tenrec} and \textbf{LastFM}-2B\footnote{http://www.cp.jku.at/datasets/LFM-2b/} \cite{LFM2B}. A detailed description can be found in Appendix A.2. We remove group-irrelevant users and items and then randomly divide interactions into training, validation, and test sets in the ratio of 7:1:2.

\begin{table*}[t]\small
\centering
\caption{Performance comparisons. Bold for the best and underline for the second best. */** indicate $p \le 0.05/0.01$ for the t-test of ITFR vs. the best baseline. $\uparrow$/$\downarrow$ means the higher/lower the better. The improvements are calculated based on the best baseline.}
\label{main}
\resizebox{0.95\textwidth}{!}{%
\begin{tabular}{l|l|ccc|cccc}
\toprule
\textbf{Dataset} & \textbf{Method}     & \textbf{P@20 $\uparrow$}                  & \textbf{R@20 $\uparrow$}                & \textbf{N@20 $\uparrow$}                    & \textbf{MIN@20 $\uparrow$}               & \textbf{CV@20 $\downarrow$}             & \textbf{UCV@20 $\downarrow$} & \textbf{ICV@20 $\downarrow$}   \\ \midrule
\multirow{8}{*}{\textbf{Movielens}}                             & \textbf{BPR}          & \underline{0.2044}                                    & \underline{0.3793}                                    & \textbf{0.3611}                                    & 0.2398                                      & 0.2026                                       & 0.0892                                        & 0.1925                                        \\
                                                       & \textbf{DPR-REO}      & 0.2013                                    & 0.3673                                    & 0.3502                                    & 0.2593                                      & 0.1567                                       & 0.0906                                        & 0.1480                                        \\
                                                       & \textbf{FairNeg}      & 0.2034                                    & 0.3761                                    & 0.359                                    & 0.2626                                      & 0.1302                                       & 0.0953                                        & 0.1214                                        \\
                                                       & \textbf{StreamDRO}    & 0.2043                                    &  \textbf{0.3794}                                    & \underline{0.3607}                                    & 0.2389                                      & 0.204                                        & 0.0859                                        & 0.1923                                        \\
                                                       & \textbf{MultiFR}      &  0.2033                                     &       0.3768                       &      0.3594                                    &       0.2403                                  &     0.1982                                       &    0.0887                                          &   0.1876                                           \\
                                                       & \textbf{GroupDRO}     & 0.2016                                    & 0.3757                                    & 0.3558                                    & \underline{0.2737}                                      & \underline{0.1145}                                       & \underline{0.0604}                                        & \underline{0.1096}                                        \\ [0.1em] \cline{2-9} 
                                                       & \textbf{ITFR(ours)}   & \textbf{0.2074$^{*}$}(+1.4\%)                                    & 0.3790(-0.1\%)                                    & 0.3605 (-0.1\%)                                    & \textbf{0.3023$^{**}$}(+10.4\%)                                     & \textbf{0.0646$^{**}$}(-43.5\%)                                       & \textbf{0.0433$^{**}$}(-28.3\%)                                        & \textbf{0.0578$^{**}$}(-47.2\%)                                        \\ \midrule
\multirow{8}{*}{\textbf{Tenrec}}                                & \textbf{BPR}          & 0.0381                                    & 0.3493                                    & 0.175                                     & 0.2703                                       & 0.1269                                       & 0.0453                                        & 0.1194                                        \\
                                                       & \textbf{DPR-REO}      & 0.0378                                    & 0.3488                                    & 0.1744                                    & 0.2764                                      & \underline{0.1126}                                       & 0.0439                                        & 0.1048                                        \\
                                                       & \textbf{FairNeg}      & \underline{0.0385}                                    & \underline{0.3517}                                    & 0.1756                                    & \underline{0.2789}                                      & 0.1138                                       & 0.0445                                        & \underline{0.1039}                                        \\
                                                       & \textbf{StreamDRO}    & 0.0382                                    & 0.3501                                    & \underline{0.1759}                                    & 0.2721                                      & 0.1223                                       & \underline{0.0372}                                        & 0.1154                                        \\
                                                       & \textbf{MultiFR}      & 0.0376                                    & 0.3471                                    & 0.1735                                    & 0.2733                                      & 0.1180                                       & 0.0429                                        & 0.1120                                        \\
                                                       & \textbf{GroupDRO}     & 0.0378                                    & 0.3471                                    & 0.1738                                    & 0.2711                                      & 0.1342                                       & 0.0460                                        & 0.1248                                        \\ [0.1em] \cline{2-9} 
                                                       & \textbf{ITFR(ours)}   & \textbf{0.0401$^{**}$}(+4.1\%)                                    & \textbf{0.3647$^{**}$}(+3.6\%)                                    & \textbf{0.1818$^{**}$}(+3.3\%)                                    & \textbf{0.3075$^{**}$}(+10.2\%)                                      & \textbf{0.0793$^{**}$}(-29.5\%)                                       & \textbf{0.0319$^{*}$}(-14.2\%)                                        & \textbf{0.0713$^{**}$}(-31.3\%)                                        \\ \midrule
\multirow{8}{*}{\textbf{LastFM}}                                & \textbf{BPR}          & 0.1090                                    & 0.1475                                     & 0.1655                                    & 0.0538                                      & 0.3398                                       & 0.0762                                        & 0.3369                                        \\
                                                       & \textbf{DPR-REO}      & 0.1078                                    & 0.1426                                    & 0.1617                                    & 0.0615                                      & 0.2989                                       & 0.0783                                        & 0.2960                                        \\
                                                       & \textbf{FairNeg}      & \underline{0.1095}                                    & \underline{0.1485}                                    & \underline{0.1662}                                    & 0.0694                                      & 0.2896                                       & 0.0794                                        & 0.2866                                        \\
                                                       & \textbf{StreamDRO}    & 0.1092                                    & 0.1478                                    & 0.1655                                    & 0.0537                                      & 0.339                                        & 0.0704                                      & 0.3362                                        \\
                                                       & \textbf{MultiFR}      & 0.1079                      & 0.1462                      & 0.1634                      & 0.0473                        & 0.3687                         & 0.0720                          & 0.3658                          \\
                                                       & \textbf{GroupDRO}     & 0.1082                                    & 0.1466                                    & 0.1642                                    & \underline{0.0896}                                      & \underline{0.1923}                                       & \underline{0.0650}                                        & \underline{0.1851}                                        \\ [0.1em] \cline{2-9} 
                                                       & \textbf{ITFR(ours)}   & \textbf{0.1114$^{**}$}(+1.7\%)                           & \textbf{0.1577$^{**}$}(+6.1\%)                           & \textbf{0.1704$^{**}$}(+2.5\%)                           & \textbf{0.0956$^{**}$}(+6.6\%)                             & \textbf{0.1772$^{**}$}(-7.8\%)                              & \textbf{0.0648}(-0.3\%)                               & \textbf{0.1680$^{**}$}(-9.8\%)                               \\ \bottomrule
\end{tabular}%
}
\end{table*}

\subsubsection{Metrics}
For accuracy metrics, we adopt the widely used NDCG@K ($N@K$), Precision@K ($P@K$), and Recall@K ($R@K$). 

For intersectional two-sided fairness metrics, given the utility definition in Eq.(\ref{utility_definition}), let $Util$ denotes the set of all the intersectional group utilities, $Util_{i,:}$ denotes the set of all the intersectional group utilities in $i$-th user group, and similarly, $Util_{:,j}$ denotes the set of all the utilities in $j$-th item group. We use the coefficients of variation \cite{ItemFair2, ItemFair3} to measure unfairness between different groups, i.e., $CV@K = \frac{\text{std}\left(Util \right)}{\text{mean}\left(Util \right)}$, where $\text{std}(\cdot)$ is the standard deviation and $\text{mean}(\cdot)$ is the average value. We also adopt a metric $MIN@K$ to measure the worst group utility. As the worst utility may be unstable \cite{UserFair5}, the average utility of the worst 25\% groups is measured.

To evaluate single-sided fairness, we also use the coefficients of variation to measure the average unfairness of utilities at the targeted single side: $ICV@K = \frac{1}{P}\sum_{i = 1}^P\frac{\text{std}\left(Util_{i, :} \right)}{\text{mean}\left(Util_{i, :} \right)}$ and $UCV@K = \frac{1}{Q}\sum_{i = 1}^Q\frac{\text{std}\left(Util_{:, i} \right)}{\text{mean}\left(Util_{:, i} \right)}$, measuring item and user fairness, respectively.

\subsubsection{Baselines}
We compare with the following baselines:

\begin{itemize}[leftmargin=*]
\item \textbf{BPR} \cite{BPR}: The classic Bayesian personalized ranking method, which does not consider fairness.
\item \textbf{StreamDRO} \cite{UserFair10}: An advanced performance-based user fairness method using a streaming distributionally robust optimization.
\item \textbf{DPR-REO} \cite{ItemFair2}: A method for performance-based item fairness using adversarial learning.
\item \textbf{FairNeg} \cite{ItemFair3}: An advanced performance-based item fairness method using adaptive fair negative sampling. 
\item \textbf{MultiFR} \cite{TwoSide1}: An in-processing two-sided fairness method using multi-objective optimization. 
\item \textbf{GroupDRO} \cite{GroupDRO}: A reweighting method adopted to this problem as Eq.(\ref{groupdroalg}), which is not originally designed for recommendation. It is a strong baseline as it is aware of intersectional groups.
\item \textbf{ITFR (ours)}: Our proposed method which uses sharpness-aware collaborative loss balance and predicted score normalization to improve intersectional two-sided fairness.
\end{itemize}

All the above methods are applied to the ranking phase in RS, and their comparative results are shown in Section 5.2. We also compare with the following two-sided reranking methods\footnote{As these two-sided reranking methods are applied in the reranking stage without conflict with our method, we evaluate the compatibility of our method with these methods in Section 5.5. \cite{TwoSide3} is excluded due to its limited applicability to two-group settings while we are handling multi-group settings.}:

\begin{itemize}[leftmargin=*]
\item \textbf{TFROM} \cite{TwoSide4}: A reranking method to improve exposure fairness for items and balance the performance losses for users.
\item \textbf{PCT} \cite{TwoSide7}: An advanced reranking method to improve exposure fairness for items and reduce exposure miscalibration for users.
\end{itemize}

\subsubsection{Implement Details}
Due to the limited space, more implementation details can be found in Appendix A.3.\footnote{Codes and data can be found at https://github.com/yfwang2021/ITFR.}

\subsection{Overall Performance}
As shown in Table \ref{main}, our proposed method ITFR significantly enhances the intersectional two-sided fairness compared to all the baselines, while also maintaining a comparable or even better accuracy. There are some further observations:
(i) Firstly, current single-sided fairness methods indeed improve their respective targeted single-sided fairness, even when using more fine-grained fairness metrics, which enables them to alleviate intersectional two-sided unfairness partially.
However, it is worth noting that they occasionally exhibit fairness compromises on the other side (e.g., Movielens), a phenomenon discussed in previous work \cite{TwoSide4}. 
(ii) Secondly, the two-sided fairness method MultiFR also partially mitigates intersectional two-sided unfairness. Nevertheless, it exhibits instability and does not consistently improve fairness across all datasets, primarily due to its lack of consideration for intersectional groups and its primary focus on optimizing item exposure fairness rather than performance-based fairness.
(iii) Thirdly, GroupDRO achieves notably higher fairness compared to the aforementioned fairness methods on the Movielens and LastFM datasets. However, its performance is less satisfactory on the Tenrec dataset. The former can be attributed to its explicit consideration of intersectional groups. The latter observation suggests that focusing solely on loss balance does not necessarily guarantee fairness, as discussed in Section 4.4.
(iv) Fourthly, our proposed method consistently outperforms all others in terms of fairness across all datasets, which demonstrates its effectiveness. 
This superior performance is attributed to its consideration of intersectional groups and its simultaneous pursuit of both two fairness goals. Furthermore, our method incurs only a negligible loss of accuracy on the Movielens dataset and even attains the best accuracy on the Tenrec and LastFM datasets.
To conclude, these results validate that our method ITFR can effectively mitigate intersectional two-sided unfairness while maintaining similar or even better accuracy.

We make further detailed analyses in the following subsections. Due to page limits, some analyses can be found in Appendix A.4, including hyperparameter analysis, case study, and compatibility with multiple single-sided attributes, and so on.

\subsection{Ablation Study}
\subsubsection{Ablation for three components}
We conduct ablation studies to assess the effectiveness of the three components within our method. Specifically,  three variants are examined: ITFR w/o sharpness-aware loss (\textbf{SA}), ITFR w/o collaborative loss balance (\textbf{CB}) and ITFR w/o predicted score normalization (\textbf{PN}). 

\begin{figure}[h]
  \centering
  \includegraphics[width=8.5cm]{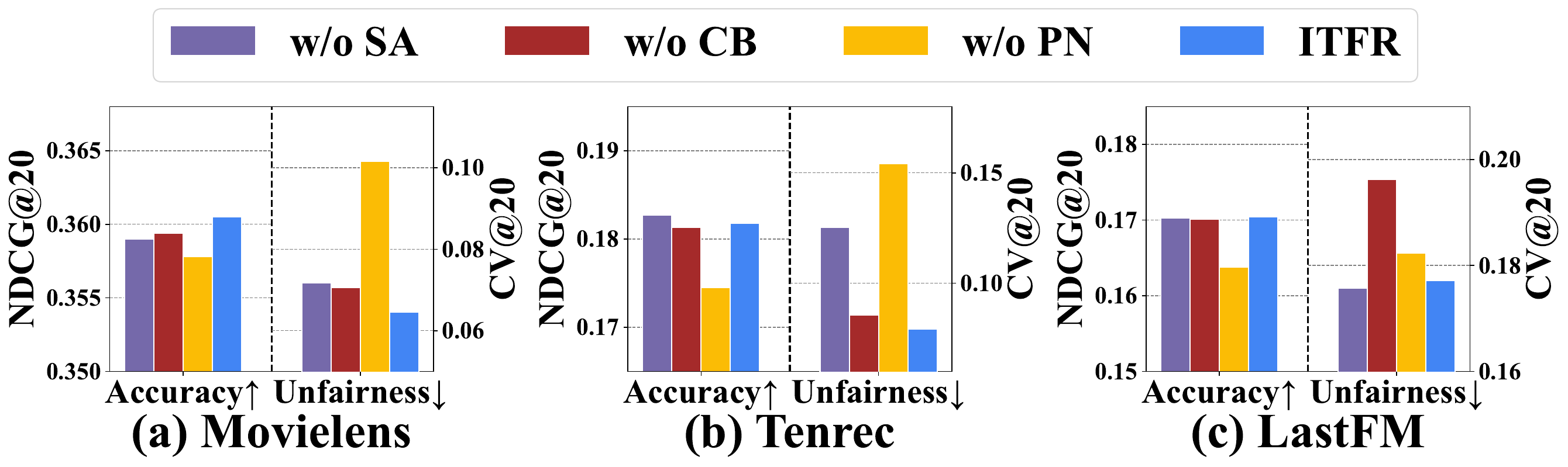}
  \caption{Ablation for three components in our method.}
  \label{ablation}
\end{figure}

As shown in Fig.\ref{ablation}, each module demonstrates efficacy in enhancing fairness. The effectiveness of these modules exhibits some variation contingent upon dataset characteristics. For instance, SA noticeably enhances fairness in the Tenrec dataset, CB is particularly effective in the LastFM dataset, and PN exhibits substantial effectiveness in both the Movielens and Tenrec datasets. As for accuracy, we observe a consistent enhancement across all three datasets with the inclusion of PN, which may be owed to its ability to alleviate popularity bias as theoretically analyzed in \cite{adaptau}.

\subsubsection{Ablation for two goals}
We further conduct ablation studies to assess the effectiveness of two fairness goals: consistent distinguishing ability (Goal 1) and treating positives fairly (Goal 2). In addition to BPR (None) and ITFR (Goal 1 \& Goal 2), we explore two variants: ITFR w/o PN (Goal 1) and BPR w. PN (Goal 2). 

\begin{figure}[h]
  \centering
  \includegraphics[width=8.5cm]{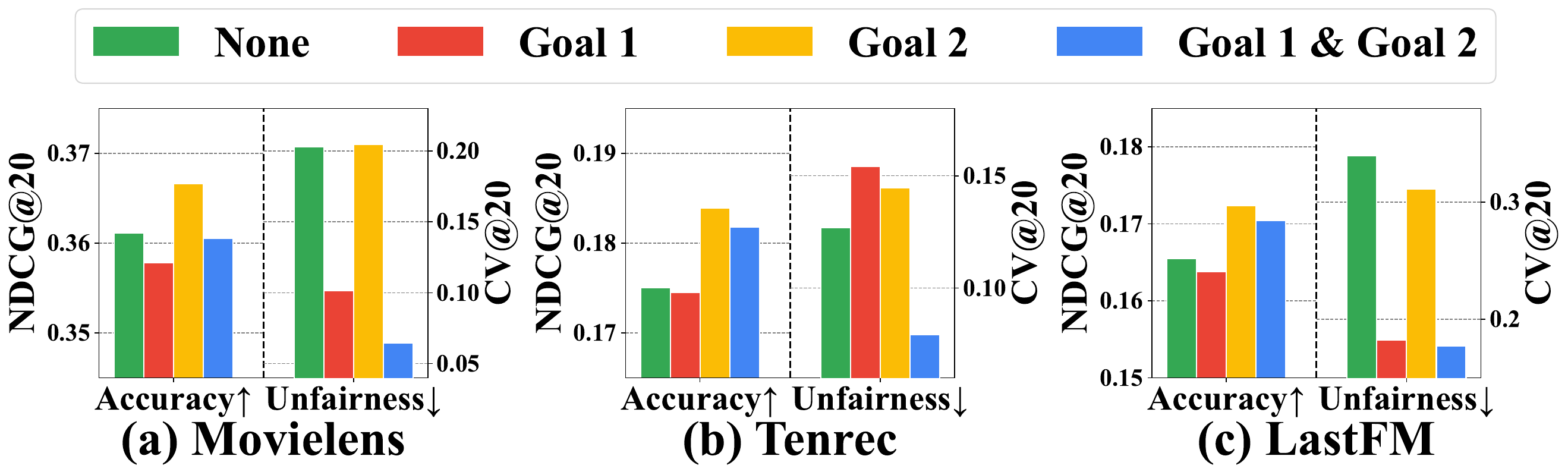}
  \caption{Ablation for two goals in our method.} 
  \label{ablation2}
\end{figure}

The results are depicted in Fig.\ref{ablation2}. We can find that both goals are important for intersectional two-sided fairness, particularly in the Tenrec dataset, where neither goal in isolation can improve fairness. Besides, Goal 2 alone yields the best accuracy, but it is not effective in improving fairness. Moreover, it enhances overall fairness less effectively than Goal 1, possibly due to the indirect way of improving Goal 2. We leave the exploration of more effective methods to achieve Goal 2 for future work.

\subsection{Compatibility with Reranking Algorithms}
Most current two-sided fairness methods \cite{TwoSide3, TwoSide4, TwoSide7} are reranking-based and focus on exposure fairness for items. Since our method is applied to the ranking phase without conflicting with these reranking methods, we next verify its compatibility with these methods. We consider two advanced two-sided fairness-aware reranking methods: TFROM \cite{TwoSide4} and PCT \cite{TwoSide7}. Exposure fairness requires a definition of a fair exposure distribution. We follow the previous work \cite{TwoSide7} that each item group should have equal exposure. For metrics, we utilize the KL-divergence \cite{ItemFair6, Joint1} between the fair distribution and the system distribution to measure item exposure unfairness, denoted as $SystemKL$. In addition, following \cite{TwoSide7}, miscalibration ($UserKL$) is used to measure whether users receive recommendations that fairly reflect their historical interests, which can be regarded as exposure fairness at the user level.

\begin{figure}[h]
  \centering
  \includegraphics[width=8.5cm]{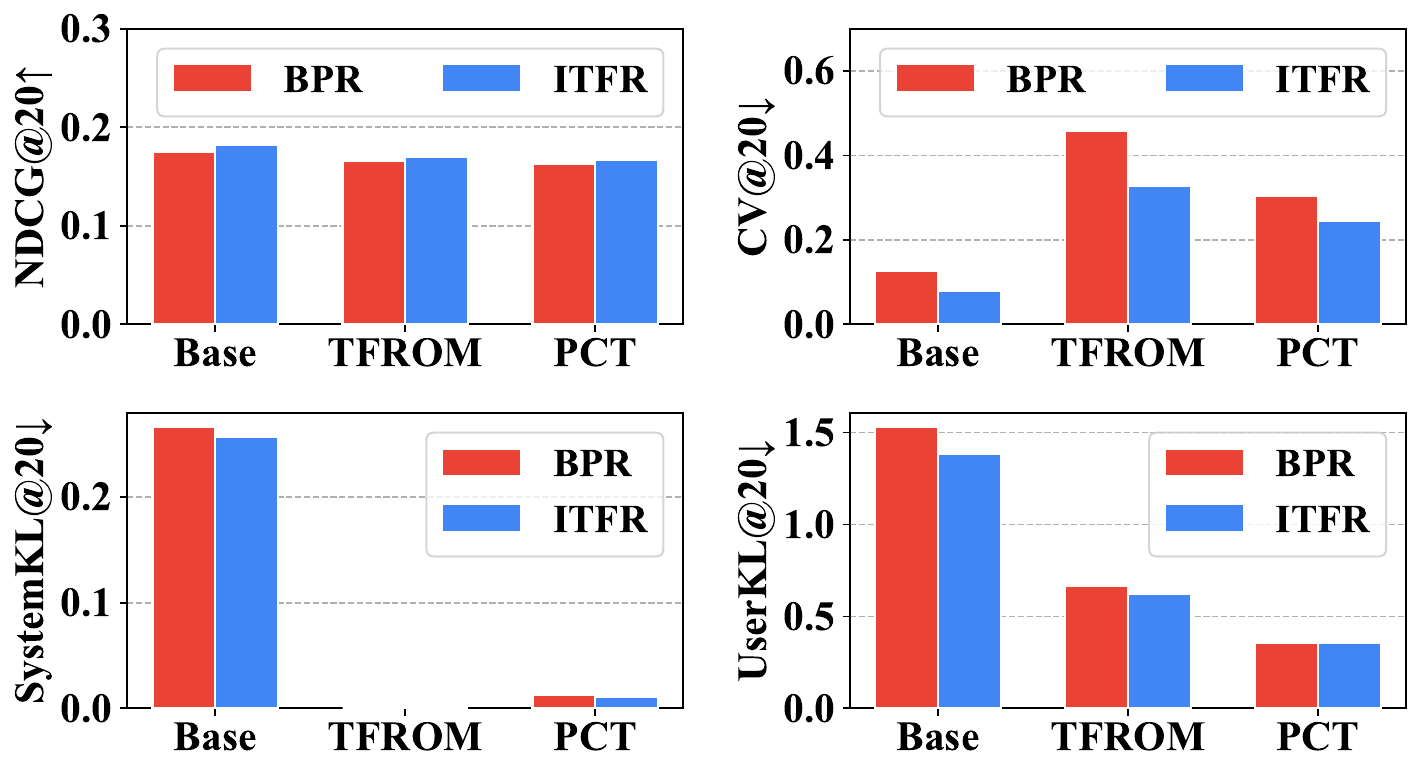}
  \caption{Reranking results on the Tenrec dataset. Results on other datasets are similar and omitted. All metrics are the lower the better except for NDCG@20.}
  \label{reranking_study}
\end{figure}

As depicted in Fig.\ref{reranking_study}, the results indicate that our method is compatible with these reranking methods. For exposure fairness, utilizing ITFR as inputs can achieve similar or even better fairness, which shows that our method does not disrupt the efficacy of these reranking methods. In addition, for intersectional two-sided fairness, it can be found that reranking has a notable detrimental effect on fairness, but ITFR still exhibits an improvement compared to BPR, underscoring its utility even in the presence of disturbances during the reranking phase.

\section{conclusions and future work}
This paper aims to mitigate the intersectional two-sided unfairness in Top-N recommendation, which current fairness methods may overlook. We first formally define the intersectional two-sided fairness and conduct empirical experiments to demonstrate the existence of such unfairness and inadequacy of current fairness methods in addressing this problem.
To fill this gap, we propose a novel method ITFR, which consists of \textit{sharpness-aware disadvantage group discovery}, \textit{collaborative loss balance}, and \textit{predicted score normalization}. 
Extensive experiments on three public datasets show that ITFR effectively alleviates the intersectional two-sided unfairness. Further experiments show that our method is also compatible with fairness-aware reranking methods. Additionally, to the best of our knowledge, our method is also the first to improve performance-based fairness for both user and item sides.

For future work, we are interested in exploring such intersectional two-sided unfairness at the individual level and exploring better ways to improve the compatibility of fairness methods in the ranking and reranking phases.

%%
%% The acknowledgments section is defined using the "acks" environment
%% (and NOT an unnumbered section). This ensures the proper
%% identification of the section in the article metadata, and the
%% consistent spelling of the heading.
% \begin{acks}
% This work is supported by the Natural Science Foundation of China (Grant No.U21B2026, 62372260), the fellowship of China Postdoctoral Science Foundation (No.2022TQ0178) and Kuaishou.
% \end{acks}

%%
%% The next two lines define the bibliography style to be used, and
%% the bibliography file.
\bibliographystyle{ACM-Reference-Format}
\bibliography{sample-base}

%%
%% If your work has an appendix, this is the place to put it.
\appendix
\section{Appendix}
% In Section A.1, we show the algorithm of sharpness-aware collaborative loss balance. In Section A.2, we introduce the datasets and corresponding statistics. Implement details can be found in Section A.3. In Section A.4, we provide the supplementary experiments.

\subsection{Learning Algorithm of Sharpness-aware Collaborative Loss Balance}
Algorithm \ref{alg1} shows the algorithm of sharpness-aware collaborative loss balance.

\begin{algorithm}
\renewcommand{\algorithmicrequire}{\textbf{Input:}}
\renewcommand{\algorithmicensure}{\textbf{Output:}}
\caption{Sharpness-aware Collaborative Loss Balance}
\label{alg1}
\begin{algorithmic}[1]
        \REQUIRE training data $\mathcal{D}$, number of intersectional groups $P \times Q$, learning rate $lr$, hyperparameters $\eta, \gamma, \rho$
        \ENSURE recommendation model $f(\theta)$
    \STATE initialize recommendation model $f(\theta)$ and group weights $w_{p,q} = \frac{1}{P \times Q}$ for $p = 1,...,P$ and for $q = 1,...,Q$.
        \FOR{$t=1$ to $T_{epoch}$}
            \FOR{batch data $\mathcal{D}_{batch}$ in $\mathcal{D}$}
                \FOR{$p=1$ to $P$}
                    \FOR{$q=1$ to $Q$}
                        \STATE calculate gradients $\nabla_\theta \mathcal{L}_{p,q}(\mathcal{D};\theta)$ of $g_{p,q}$ 
                        \STATE calculate sharpness-aware loss $\hat{\mathcal{L}}_{p,q}$ based on Eq.(\ref{sharpness_cal})
                        \STATE calculate sharpness-aware gradients $\nabla_\theta \hat{\mathcal{L}}_{p,q}$
                    \ENDFOR
                \ENDFOR

                % \STATE calculate group weight $\beta_{p,q}$ for any $p,q$ based on Eq.(\ref{all_contrib})
                % \STATE calculate $\mathcal{C}(g_{p,q} \rightarrow g_{a,b})$ for any $p,q,a,b$ (Eq.(\ref{pair_contrib}))
                \STATE calculate $\mathcal{C}(g_{p,q})$ for any $p,q$ based on Eq.(\ref{all_contrib})
                \STATE update group weights $w_{p,q}$ for any $p,q$ based on Eq.(\ref{w_update})
                \STATE update $\theta \leftarrow \theta - lr* \left(\sum_{p,q} w_{p,q} \cdot \nabla_\theta \hat{\mathcal{L}}_{p,q}\right)$
            \ENDFOR
        \ENDFOR
        \RETURN recommendation model $f(\theta)$
\end{algorithmic}  
\end{algorithm}

\subsection{Datasets and Statistics}
We conduct experiments on three public datasets below. The statistics of the processed datasets are shown in Table \ref{datasets}.

\textbf{Movielens-1M.} This dataset contains 1 million movie ratings with user and item profiles. 
Gender is used to divide user groups, while movie genres are utilized to divide item groups, a commonly used group division in fairness studies \cite{ItemFair2, TwoSide1, ItemFair3}. Specifically, we select six genres (`Sci-Fi', `Adventure', `Crime', `Romance', `Children's', `Horror') as previously used in \cite{ItemFair2}. 

\textbf{Tenrec-QBA.} This dataset is collected from a news recommendation platform comprising 348K article clicks. The age is used to divide user groups. Specifically, as the age attribute is grouped in decades with a disrupted order and some decades have little data, we choose the three most popular attribute values (`1', `7', `8'). For items, we use the article channel to divide groups and select the four most popular attribute values (`104', `113', `124', `127').

\textbf{LastFM-2B.} This dataset contains two billion listening events, some of which include genre information. Users are grouped based on gender. For items, we choose four of the most popular genres with large style differences: (`rock', `pop', `jazz', `ambient'). We removed users/items without group value in the preprocessing.

\begin{table}[h]\small
\centering
\caption{Statistics of the processed datasets.}
\label{datasets}
\begin{tabular}
{lllll} 
%\hline
\toprule
\textbf{Dataset} & \textbf{\#Users} & \textbf{\#Items} & \textbf{\#Interactions} & \textbf{Density}  \\ 
\midrule
\textbf{Movielens} & 5,977 & 1,200 & 396,207 & 0.0552 \\
\textbf{Tenrec} & 11,376 & 1,015 & 132,981 & 0.0115 \\
\textbf{LastFM} & 20,847 & 18,625 & 1,785,420 & 0.0046 \\
\bottomrule
%\hline
\end{tabular}%}
\end{table}

We also provide the interaction distributions for the three datasets in the following Table \ref{id:ml}, \ref{id:tenrec} and \ref{id:lastfm}. Note that, in each of the three datasets involved in our experiments, the percentages of the sparsest intersectional groups are as follows: (2.0\%, Movielens: Female-Adventure), (1.5\%, Tenrec: 6-124), and (1.3\%, LastFM: Female-Jazz). It indicates well that even for such a significant imbalance, our method demonstrates effective results on all three datasets.

\begin{table}[h]
\centering
\caption{Distribution of interactions on the Movielens dataset.}
\label{id:ml}
\resizebox{0.47\textwidth}{!}{%
\begin{tabular}{c|cccccc}
\toprule
\textbf{User\textbackslash{}Item} & \textbf{Sci-Fi} & \textbf{Adventure} & \textbf{Children's} & \textbf{Crime} & \textbf{Horror} & \textbf{Romance} \\ \midrule
\textbf{Male}                     & 15.7\%          & 8.3\%              & 8.2\%               & 12.2\%         & 10.2\%          & 19.3\%           \\
\textbf{Female}                   & 3.2\%           & 2.0\%              & 3.6\%               & 3.3\%          & 2.5\%           & 10.8\%           \\ \bottomrule
\end{tabular}%
}
\end{table}
\begin{table}[h]
\centering
\caption{Distribution of interactions on the Tenrec dataset.}
\label{id:tenrec}
\resizebox{0.28\textwidth}{!}{%
\begin{tabular}{c|cccc}
\toprule
\textbf{User\textbackslash{}Item} & \textbf{104} & \textbf{113} & \textbf{124} & \textbf{127} \\ \midrule
\textbf{5}               & 15.1\%       & 13.2\%       & 4.3\%        & 6.8\%        \\
\textbf{6}               & 5.1\%        & 4.7\%        & 1.5\%        & 2.5\%        \\
\textbf{7}               & 13.0\%       & 20.2\%       & 3.7\%        & 9.4\%        \\ \bottomrule
\end{tabular}%
}
\end{table}
\begin{table}[h]
\centering
\caption{Distribution of interactions on the Last.FM dataset.}
\label{id:lastfm}
\resizebox{0.3\textwidth}{!}{%
\begin{tabular}{c|cccc}
\toprule
\textbf{User\textbackslash{}Item} & \textbf{Rock} & \textbf{Pop} & \textbf{Jazz} & \textbf{Ambient} \\ \midrule
\textbf{Male}                     & 38.7\%        & 23.1\%       & 7.9\%         & 9.9\%            \\
\textbf{Female}                   & 8.8\%         & 8.4\%        & 1.3\%         & 1.5\%            \\ \bottomrule
\end{tabular}%
}
\end{table}

\subsection{Implement Details}
We use the classic MF \cite{MF} as user and item encoders for all the methods. The embedding size is set to 64 for all the methods. Adam \cite{adam} is used as the optimizer. The learning rate is set to 1e-3 with the L2 regularization tuned in [0, 1e-7, 1e-6, 1e-5, 1e-4]. The batch size is set to 1024. For fair comparisons, uniform negative sampling is applied to all the models during training, except for FairNeg mixed with a fair negative distribution. The negative sampling number for training is set to 1. The additional hyperparameters for the baselines are fine-tuned following their original paper. The best models are selected based on the performance of the validation set within 200 epochs. We repeat each experiment 5 times and report the average results, and perform statistical tests (i.e., t-test).

\subsection{Further Analyses}
\subsubsection{Hyperparameter Analysis}
\begin{figure}[h]
  \centering
  \includegraphics[width=8.5cm]{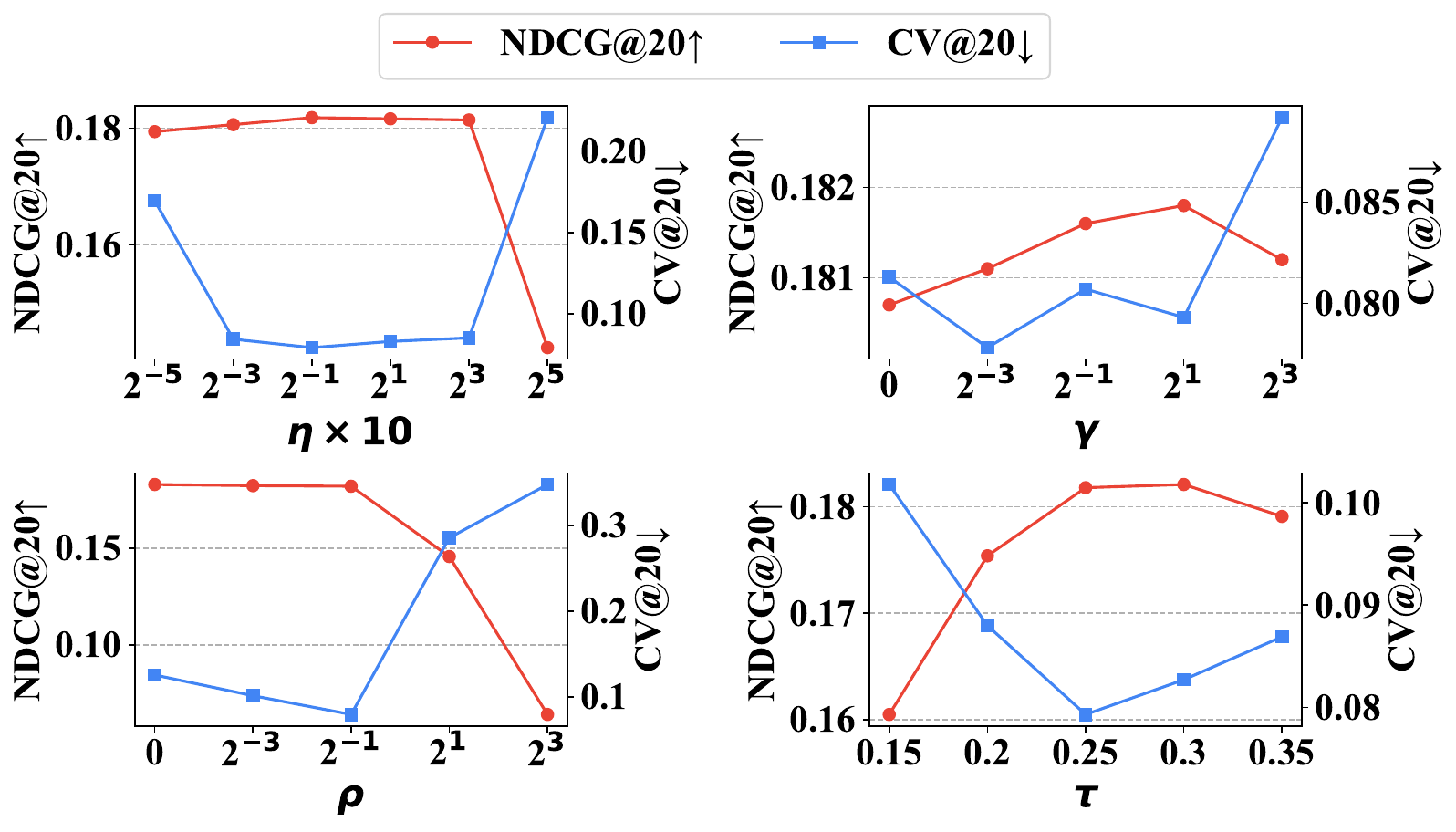}
  \caption{Hyperparameter analysis on the Tenrec dataset.}
  \label{hyper_all}
\end{figure}

We further analyze the influence of the involved hyperparameters, specifically $\rho$ in SA, $\eta$ and $\gamma$ in CB, and $\tau$ in PN. The results are presented in Fig.\ref{hyper_all}. For all parameters, fairness tends to initially improve and subsequently decline as the parameter value increases. A similar pattern is observed in accuracy, though with some variations in magnitude. For $\rho, \eta, \gamma$, increasing the value within a specific range amplifies the impact of the corresponding component, resulting in further improvements in fairness. However, excessively large values can compromise optimization stability and result in a rapid decline in performance. Regarding $\tau$, it influences the gradient magnitude at each update to some extent; thus, too large or too small values are not suitable.

\subsubsection{Case Study}
We conduct a case study to show the improvement of certain intersectional groups. The results are shown in Table \ref{case:bpr} and \ref{case:itfr}. Firstly, we observe that ITFR effectively reduces unfairness among intersectional groups, as evidenced by the overall $CV@20$ decreasing from 0.2026 to 0.0646. Additionally, the utility of the worst group (Female-Horror) improves from 0.2145 to 0.2958. Furthermore, for each of the five worst intersectional groups (i.e., Female-Horror, Male-Horror, Male-Children's, Male-Romance, Female-Adventure), ITFR effectively enhances their utilities, which validates the effectiveness of ITFR.

\begin{table}[h]
\centering
\caption{Results of BPR on the Movielens dataset. The worst 5 groups of BPR are underlined.}
\label{case:bpr}
\resizebox{0.47\textwidth}{!}{%
\begin{tabular}{c|cccccc}
\toprule
\textbf{User\textbackslash{}Item} & \textbf{Sci-Fi} & \textbf{Adventure} & \textbf{Children's} & \textbf{Crime} & \textbf{Horror} & \textbf{Romance} \\ \midrule
\textbf{Male}                     & 0.4520          & 0.3591             & \underline{0.2578}              & 0.3982         & \underline{0.2471}          & \underline{0.2668}           \\
\textbf{Female}                   & 0.3422          & \underline{0.3079}             & 0.3222              & 0.3577         & \underline{0.2145}          & 0.3165           \\ \bottomrule
\end{tabular}%
}
\end{table}

\begin{table}[h]
\centering
\caption{Results of ITFR(ours) on the Movielens dataset. The worst 5 groups of BPR are underlined, and the ↑/↓ means better or worse recommendations compared to BPR.}
\label{case:itfr}
\resizebox{0.47\textwidth}{!}{%
\begin{tabular}{c|cccccc}
\toprule
\textbf{User\textbackslash{}Item} & \textbf{Sci-Fi} & \textbf{Adventure} & \textbf{Children's} & \textbf{Crime} & \textbf{Horror} & \textbf{Romance} \\ \midrule
\textbf{Male}                     & 0.3539↓         & 0.3520↓            & \underline{0.3068}↑             & 0.3636↓        & \underline{0.3270}↑         & \underline{0.3043}↑          \\
\textbf{Female}                   & 0.3145↓         & \underline{0.3156}↑            & 0.3356↑             & 0.3377↓        & \underline{0.2958}↑         & 0.3136↓          \\ \bottomrule
\end{tabular}%
}
\end{table}

\subsubsection{Compatibility with Multiple Single-sided Attributes}
In the experimental section, users and items are divided by a single attribute for simplicity. ITFR can also be applied in the case of the presence of multiple attributes on a single side. As discussed in Section 2.3, we can use the Cartesian product of multiple single-sided attributes as the group division. As shown in Table \ref{multiattri}, we present results on the Movielens dataset, where users are divided by both gender and age (then there are $2\times3$ user groups in total, namely, Male-Below25, Female-Below25, Male-Above45, Female-Above45, Male-Middle, and Female-Middle). It can be found that our method remains effective in mitigating intersectional unfairness while maintaining similar accuracy, which validates the compatibility with multiple single-sided attributes of our method.

\begin{table}[h]
\centering
\caption{Results with multiple single-sided attributes on the Movielens dataset. Better results are in bold.}
\label{multiattri}
\resizebox{0.47\textwidth}{!}{%
\begin{tabular}{c|ccccccc}
\toprule
\textbf{}           & \textbf{P@20↑}  & \textbf{R@20↑}  & \textbf{N@20↑}  & \textbf{MIN@20↑} & \textbf{CV@20↓} & \textbf{UCV@20↓} & \textbf{ICV@20↓} \\ \midrule
\textbf{BPR}        & 0.2044          & \textbf{0.3793} & \textbf{0.3611} & 0.2350           & 0.2169          & 0.1253           & 0.2067           \\
\textbf{ITFR(ours)} & \textbf{0.2073} & 0.3786          & 0.3608          & \textbf{0.2876}  & \textbf{0.1019} & \textbf{0.0770}  & \textbf{0.0938}  \\ \bottomrule
\end{tabular}%
}
\end{table}

\subsubsection{Other Fairness Metrics}
In the experimental section, we use $UCV@20$ and $ICV@20$ to measure single-sided fairness, which computes the average of the single-sided unfairness (measured by coefficients of variation) at each of the values taken on the other side, allowing us to capture more localized single-sided unfairness. We believe that the different utilities of users on specific item groups are also part of the user unfairness and should not be neutralized by what happens on other item groups. The same is true for item unfairness. Similar fine-grained metrics are used by \cite{UserFair9}, but they are only applied to rating prediction and the two-group setting.

\begin{figure}[h]
  \centering
  \includegraphics[width=8.5cm]{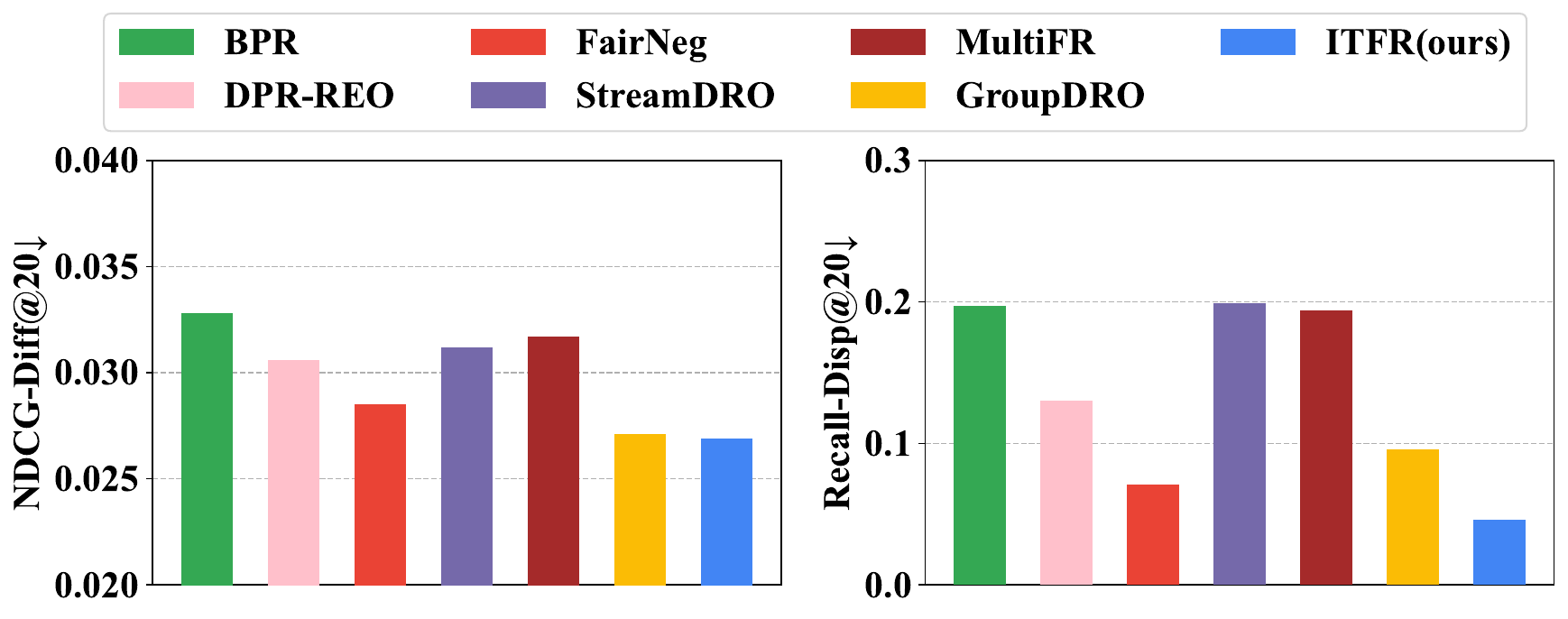}
  \caption{Results of other fairness metrics on the Movielens dataset. Both metrics are the lower the better.}
  \label{other_metrics}
\end{figure}

Our method also performs well in other common metrics. We present results on the Movielens dataset as shown in Fig.\ref{other_metrics}, using the user fairness metric \textit{NDCG-Diff} (the recommendation performance disparity between two user groups \cite{UserFair6,UserFair10}) and the item fairness metric \textit{Recall-Disp} (the coefficients of variation of the recall at the item group level \cite{ItemFair2,ItemFair3}). It can be found that our method still achieves the best fair results.

\subsubsection{Other Recommendation Backbone}
In Section 5, we use the classic MF as the encoder of users and items. ITFR can also combine with more advanced encoders. Below we provide the results of LightGCN \cite{LightGCN} and the combination with our method ITFR on the Movielens dataset in the following Table \ref{lightgcn}. It can be found that our method is also effective in improving fairness with a stronger recommendation model. It also shows that our method is flexible enough to be easily combined with newer recommendation models.

\begin{table}[h]
\centering
\caption{Results of LightGCN combined with ITFR on the Movielens dataset. Better results are in bold.}
\label{lightgcn}
\resizebox{0.46\textwidth}{!}{%
\begin{tabular}{c|ccccccc}
\hline
\textbf{}              & \textbf{P@20↑}  & \textbf{R@20↑}  & \textbf{N@20↑}  & \textbf{MIN@20↑} & \textbf{CV@20↓} & \textbf{UCV@20↓} & \textbf{ICV@20↓} \\ \hline
\textbf{LightGCN}      & \textbf{0.2109} & \textbf{0.3913} & \textbf{0.3743} & 0.2460           & 0.1956          & 0.1057           & 0.1829           \\
\textbf{w. ITFR} & 0.2080          & 0.3899          & 0.3701          & \textbf{0.2869}  & \textbf{0.1036} & \textbf{0.0526}  & \textbf{0.0999}  \\ \hline
\end{tabular}%
}
\end{table}

\subsubsection{Combination of Two Kinds of Single-sided Groups}
To the best of our knowledge, there is no performance-based two-sided fairness method. One might be curious about how well the performance-based two-sided fairness method performs in terms of intersectional two-sided fairness. To investigate this, we adopt GroupDRO in a two-sided way. Specifically, we divide different single-sided groups by the user side, as well as different single-sided groups by the item side, and then assign different weights to these single-sided groups in the GroupDRO manner, denoted as ``Two-sided''. In contrast, the GroupDRO in an intersectional two-sided way as introduced in Section 5.1 is denoted as ``Intersectional''. 

\begin{figure}[h]
  \centering
  \includegraphics[width=8.5cm]{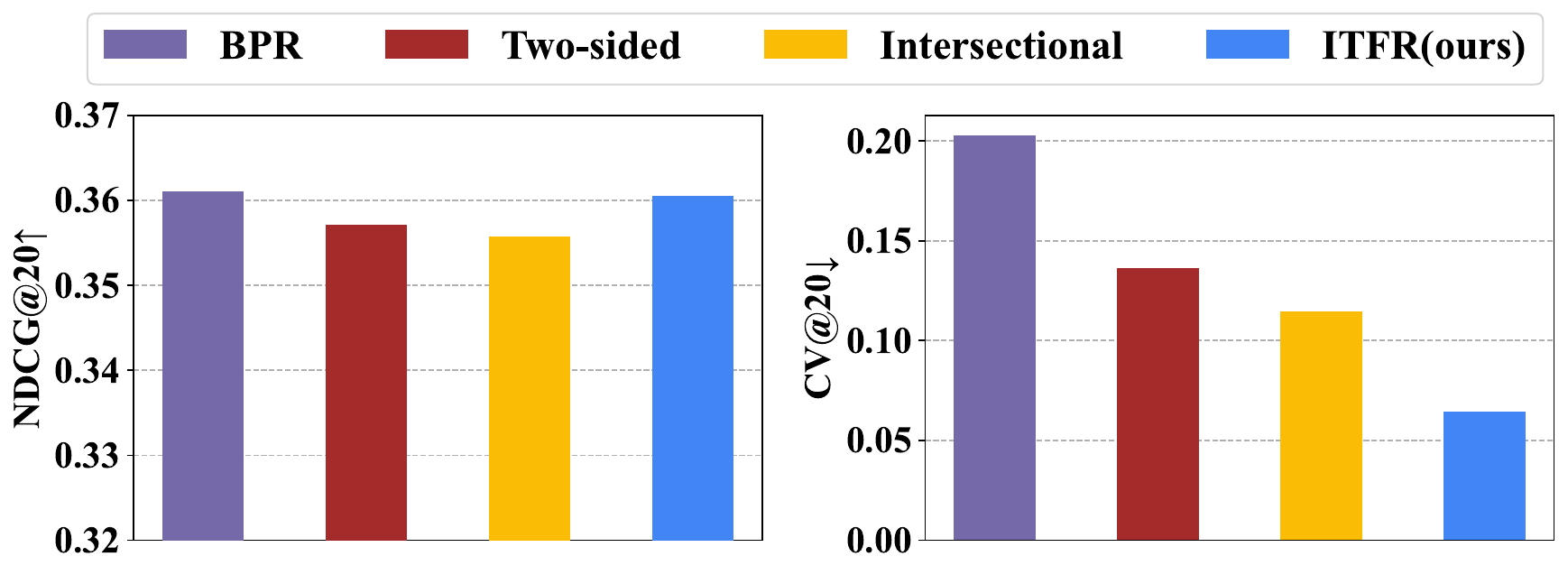}
  \caption{Comparison with GroupDRO combined with different strategies on the Movielens dataset.}
  \label{twosided}
\end{figure}

The results are presented in Fig.\ref{twosided}. It can be found that optimizing the combination of two kinds of single-sided groups indeed mitigates intersectional two-sided unfairness to some extent, but optimizing directly for the intersectional two-sided groups (i.e., ``Intersectional'') achieves better fairness. This is because the latter can handle the diagonal situation in the illustration in Fig.\ref{illustration}, which cannot be identified by the former. Besides, our method significantly outperforms this strategy in terms of both accuracy and fairness.

\end{document}